\documentclass{article}
\usepackage{amsmath,amssymb}

\begin{document}

\title{Solutions for a class of fifth-order nonlinear
partial differential system\thanks{%
2000 Mathematics Subject Classification: 35C05.}}
\date{{\small 06 Oct. 2008}}

\author{Cesar A Gomez S.
\thanks{%
Department of Mathematics, Universidad Nacional de Colombia,
Bogot\'a, Colombia  \emph{e-mail}: cagomezsi@unal.edu.co}
}

\maketitle

\begin{abstract}
In this paper we use the generalized tanh method to obtain exact solutions
for a class of fifth-order nonlinear systems. A particular case is given by the
integrable Mikhailov-Novikov-Wang system (MNW).  Periodic and
solitons solutions are formally derived . The \emph{Mathematica} is used.
\end{abstract}

\section{Introduction}

The class of fifth-order systems that we consider reads
\begin{equation}
\begin{cases}
u_t+\rho u_{xxxxx}+\gamma uu_{xxx}+\beta u_xu_{xx}+\alpha
u^2u_x-w_x=0\\
w_t+6wu_{xxx}+2u_{xx}w_x-96wuu_x-16w_xu^2=0,
\end{cases}
\end{equation}
where $\alpha$, $\beta$, $\gamma$ and $\rho$ are arbitrary nonzero
and real parameters, and  $u=u(x,t)$, $w=w(x,t)$ are
differentiable functions. Lost of forms of (1) can be
constructed by changing the values of parameters.  In the particular
case $\alpha=-80$, $\beta=50$, $\gamma=20$, $\rho=-1$, (1)
reduces to the new Mikhailov-Novikov-Wang system [1][2][3][4][5] (MVW)
\begin{equation}
\begin{cases}
u_t= u_{xxxxx}-20 uu_{xxx}-50 u_xu_{xx}+80u^2u_x+w_x\\
w_t=-6wu_{xxx}-2u_{xx}w_x+96wuu_x+16w_xu^2,
\end{cases}
\end{equation}
which was  derived by the authors in [2] using the
symmetry approach. We refer to [2] for more details about this
system . In the case that $w(x,t)=0$ and $\rho=1$, the system (1)
reduces to fifth-order KdV equation [6][7][8][9][10]
\begin{equation}
u_t+u_{xxxxx}+\gamma uu_{xxx}+\beta u_xu_{xx}+\alpha u^2u_x=0.
\end{equation}
This equation has been recently studied in [7][8][9][10]. Some important
particular cases of (3) are given by Lax, Sawada--Kotera,
Kaup--Kupershdmit, and Ito equations [7][8][9][10].\par The searching of exact solutions of
nonlinear partial differential equations is the great importance
for many researches. A variety of powerful methods such that
\text{tanh method} and \text{generalized tanh method} [6][7][11][12], \newline  \text{general projective Riccati
equations method}[13][14] [15] and other methods (see for instance [7][8][9][10][16][17]) have been developed in
this direction.  In this work, we will use
the generalized tanh method [8][11][12] to construct  periodic and
solitons solutions to (1). This paper is organized as follows: In Sec. 1,
we will be reviewed briefly of the generalized tanh method. In
Sec. 2, we give the mathematical framework to search exact
solutions to (1). In sec. 3, we obtain exact solutions to (1).  Finally some conclusions are given.

\section{The generalized tanh method }
The generalized tanh method can be summarized as follows. For a
given nonlinear equation that does not explicitly involve
independent variables
\begin{equation}
P(u,u_x,u_t,u_{xx},u_{xt},u_{tt},\ldots)=0,
\end{equation}
we use the wave transformation
\begin{equation}
u(x,t)=v(\xi),\;\;\;\xi=x+\lambda t,
\end{equation}
where $\lambda$  is a constant.
\par Under the transformation (5), (4) reduces to ODE in the function $v(\xi)$
\begin{equation}
P(v,v',v'',\ldots)=0.
\end{equation}
 The next crucial step is to introduce a new variable
 $\phi(\xi)$ which is a solution of the Riccati equation
\begin{equation}
\phi'(\xi)=\phi(\xi)^2+k,
\end{equation}
where $k$ is a constant.\par It is well-known that the solutions of the
equation (7) are given by
\begin{equation}
\phi(\xi)=
\begin{cases}
-\frac{1}{\xi}, & k=0\\
\sqrt{k}\tan(\sqrt{k}\xi) & k>0\\
-\sqrt{k}\cot(\sqrt{k}\xi) & k>0\\
-\sqrt{-k}\tanh(\sqrt{-k}\xi) & k<0\\
-\sqrt{-k}\coth(\sqrt{-k}\xi) & k<0.
\end{cases}
\end{equation}
We seek solutions to (6) in the form
\begin{equation}
\sum^{M}_{i=0}a_i\phi(\xi)^{i},
\end{equation}
where $\phi(\xi)$ satisfies the Riccati equation (7), and $a_i$
are unknown constants. The integer $M$ can be determined by
balancing the highest derivative term with nonlinear terms in (6),
before the $a_i$ can be computed. Substituting (9) along with (7)
into (6) and collecting all terms with the same power
$\phi(\xi)^{i}$, we get a polynomial in the variable $\phi(\xi)$.
Equaling the coefficients of this polynomial to zero, we can
obtain a system of algebraic equations, from which  the constants
$a_i$, $\lambda$ ($i=1,2,\ldots,M$) are obtained explicitly.
Lastly, we found solutions for (4) in the original variables.

\section{ Exact Solutions to (1).}

 To search exact solutions of the system
(1), we use the traveling wave transformation
\begin{equation}
\begin{cases}
u(x,t)=v(\xi)\\
w(x,t)=w(\xi)\\
\xi=x+\lambda t.
\end{cases}
\end{equation}
Under the transformation (10), (1) reduces to following nonlinear
ordinary differential equations system with constant coefficients
\begin{equation}
\begin{cases}
\lambda
v'(\xi)+\rho v^{(5)}(\xi)+\gamma v(\xi)v^{(3)}(\xi)+\beta v'(\xi)v''(\xi)+\alpha v^2(\xi)v'(\xi)-w'(\xi)=0\\
\lambda
w'(\xi)+6w(\xi)v'''(\xi)+2v''(\xi)w'(\xi)-96w(\xi)v(\xi)v'(\xi)-16w'(\xi)v^2(\xi)=0.
\end{cases}
\end{equation}
From the first equation in (11) we obtain
\begin{equation}
w'(\xi)=\lambda v'(\xi)+\rho v^{(5)}(\xi)+\gamma
v(\xi)v^{(3)}(\xi)+\beta v'(\xi)v''(\xi)+\gamma v^2(\xi)v'(\xi),
\end{equation}
which can be written as
\begin{equation}
(\lambda
v(\xi)+\frac{\alpha}{3}v^3(\xi)+\frac{\beta-\gamma}{2}(v'(\xi))^2+\gamma
v(\xi)v''(\xi)+\rho v^{(4)}(\xi)-w(\xi))'=0.
\end{equation}
Integrating (13) once with respect to $\xi$ we obtain
\begin{equation}
\lambda
v(\xi)+\frac{\alpha}{3}v^3(\xi)+\frac{\beta-\gamma}{2}(v'(\xi))^2+\gamma
v(\xi)v''(\xi)+\rho v^{(4)}(\xi)-w(\xi)=c,
\end{equation}
where $c$ is integration constant. We take $c=0$. Due to (14)
\begin{equation}
w(\xi)=\lambda
v(\xi)+\frac{\alpha}{3}v^3(\xi)+\frac{\beta-\gamma}{2}(v'(\xi))^2+\gamma
v(\xi)v''(\xi)+\rho v^{(4)}(\xi).
\end{equation}
Substituting (15) and (12) in the second equation in (11) and
after simplifications we obtain the following ordinary
differential equation
\begin{equation}
\begin{cases}
\lambda^2v'(\xi)+\lambda \rho v^{(5)}+\lambda(6+\gamma)
v(\xi)v'''(\xi)+\lambda (\beta+2)
v'(\xi)v''(\xi)+\lambda(\gamma-112)
v^2(\xi)v'(\xi)+2\rho v''(\xi)v^{(5)}(\xi)+\\
8\gamma v(\xi)v''(\xi)v'''(\xi)+2\beta v'(\xi)(v''(\xi))^2-2(47\gamma+8\beta)v^2(\xi)v'(\xi)v''(\xi)\\
-16\rho v^2(\xi)v^{(5)}+2(\alpha-8\gamma)v^3(\xi)v'''(\xi)-16(2\alpha+\gamma)v^4(\xi)v'(\xi)+6\rho v'''(\xi)v^{(4)}+3(\beta-\gamma)(v'(\xi))^2v'''(\xi)\\
-96\rho v(\xi)v'(\xi)v^{(4)}-48(\beta-\gamma)v(\xi)(v'(\xi))^3=0.
\end{cases}
\end{equation}
\newline

According to the described above method, we seek solutions to (16) as
\begin{equation}
v(\xi)=a_0+a_1\phi(\xi)+a_2(\phi(\xi))^2,
\end{equation}
where $\phi(\xi)$ satisfies (7).
Substituting (17) into (16) and using (7) we obtain a polynomial in
$\phi(\xi)$. Equaling the coefficients of this polynomial to zero,
and solving the resulting algebraic system respect to unknowns
variables $k$, $\lambda$, $a_0$, $a_1$ and $a_2$ with aid the
Mathematica  we find the following sets of solutions:

\begin{itemize}
\item
$a_1=0$, $a_2=\frac{3}{4}$, $k=\sqrt{\lambda}$,
$a_0=\frac{\sqrt{\lambda}}{2}$.
\item  .
$a_1=0$, $a_2=\frac{3}{4}$, $k=-\sqrt{\lambda}$,
$a_0=-\frac{\sqrt{\lambda}}{2}$.
\end{itemize}

Using (8) we obtain the following solutions to (16):
\newline
For $\lambda>0$:

\begin{enumerate}
\item
 $v_1(\xi)=\frac{\sqrt{\lambda}}{2}+\frac{3}{4}\sqrt{\lambda}\cot^2(\lambda^{\frac{1}{4}}\xi)$.
 \item
 $v_2(\xi)=-\frac{\sqrt{\lambda}}{2}+\frac{3}{4}\sqrt{\lambda}\coth^2(\lambda^{\frac{1}{4}}\xi)$.
 \item
 $v_3(\xi)=\frac{\sqrt{\lambda}}{2}+\frac{3}{4}\sqrt{\lambda}\tan^2(\lambda^{\frac{1}{4}}\xi)$.
 \item
 $v_4(\xi)=-\frac{\sqrt{\lambda}}{2}+\frac{3}{4}\sqrt{\lambda}\tanh^2(\lambda^{\frac{1}{4}}\xi)$.
 \end{enumerate}

 Therefore by (10) and (15) the exact solutions to
 system (1) are given by :
 \newline
 For $\lambda>0$:
\begin{enumerate}
\item
 $u_1(x,t)=\frac{\sqrt{\lambda}}{2}+\frac{3}{4}\sqrt{\lambda}\cot^2(\lambda^{\frac{1}{4}}(x+\lambda t))$\\
 $w_1(x,t)=\frac{\lambda^{\frac{3}{2}}}{192}(-48-\alpha+9\csc^2(\lambda^{\frac{1}{4}}(x+\lambda t)(\alpha+16(1+\gamma+16\rho)+3(\alpha+8(\beta+2\gamma+80\rho))\cot^2(\lambda^{\frac{1}{4}}(x+\lambda t)\csc^2(\lambda^{\frac{1}{4}}(x+\lambda
 t))))))$.
 \item
 $u_2(x,t)=-\frac{\sqrt{\lambda}}{2}+\frac{3}{4}\sqrt{\lambda}\coth^2(\lambda^{\frac{1}{4}}(x+\lambda t)).$\\
 $w_1(x,t)=\frac{\lambda^{\frac{3}{2}}}{192}(48+\alpha+9\text{sech}^2(\lambda^{\frac{1}{4}}(x+\lambda t)(\alpha+16(1+\gamma+16\rho)+3(\alpha+8(\beta+2\gamma+80\rho))\coth^2(\lambda^{\frac{1}{4}}(x+\lambda t)\text{csch}^2(\lambda^{\frac{1}{4}}(x+\lambda
 t))))))$.
 \item
 $u_3(x,t)=\frac{\sqrt{\lambda}}{2}+\frac{3}{4}\sqrt{\lambda}\tan^2(\lambda^{\frac{1}{4}}(x+\lambda t)).$\\
 $w_1(x,t)=\frac{\lambda^{\frac{3}{2}}}{192}(-48-\alpha+9\sec^2(\lambda^{\frac{1}{4}}(x+\lambda t)(\alpha+16(1+\gamma+16\rho)+3(\alpha+8(\beta+2\gamma+80\rho))\sec^2(\lambda^{\frac{1}{4}}(x+\lambda t)\tan^2(\lambda^{\frac{1}{4}}(x+\lambda
 t))))))$.
 \item
 $u_4(x,t)=-\frac{\sqrt{\lambda}}{2}+\frac{3}{4}\sqrt{\lambda}\tanh^2(\lambda^{\frac{1}{4}}(x+\lambda t)).$\\
$w_1(x,t)=\frac{\lambda^{\frac{3}{2}}}{192}(48+\alpha-9\text{sech}^2(\lambda^{\frac{1}{4}}(x+\lambda
t))(\alpha+16(1+\gamma+16\rho))+27(\alpha+8(\beta+2\gamma+80\rho))\text{sech}^4(\lambda^{\frac{1}{4}}(x+\lambda
t))-27(\alpha+8(\beta+2\gamma+80\rho))\text{sech}^6(\lambda^{\frac{1}{4}}(x+\lambda
t)))$.
 \end{enumerate}

\section{ Conclusions}
In the previous sections, we have presented an analysis over a
generalizations of the MNW system. A large number of forms of the
fifth-order system with exact solutions can be constructed. Exact
solutions for some particular cases such that
Mikhailov-Novikov-Wang system can be derived. The
generalized tanh method provides a straightforward algorithm to
compute particular periodic and solitons solutions for a large
class of nonlinear systems.

\bigskip

\noindent\textbf{Acknowledgments:} The authors wants to express
their gratitude to professor A. Sinitsyn for his helpful
suggestions and recommendations about this paper.

\end{document}